\documentclass[preprint,aps,amsmath,superscriptaddress,nofootinbib,tightenlines]{revtex4}
\usepackage{graphicx,psfrag}
\usepackage{bm}
\usepackage{epsfig}
\def\OMIT#1{}

\newcommand{\nn}{\nonumber}

\newcommand{\bn}{{\bar n}}
\newcommand{\bea}{\begin{eqnarray}}
\newcommand{\eea}{\end{eqnarray}}

\newcommand{\bnP}{\bar {\cal P}}

\newcommand{\cP}{{\cal P}}

\newcommand{\mup}{M_\Upsilon}
\newcommand{\mpsi}{M_\psi}
\newcommand{\pup}{p_\Upsilon}

\newcommand{\jpsi}{J/\psi}

\begin{document}

%%%%%%%%%%%%%%%%%%%%%%%%%%%%%%%%%%%%%%%%%%
%Some more stuff to get graphics to work
%
%\ifpdf
%\DeclareGraphicsExtensions{.pdf, .jpg,.ps}
%\newcommand{\picspace}{\vspace{-2.5in}}
%\newcommand{\picspacehalf}{\vspace{-1.75in}}
%\else
%\DeclareGraphicsExtensions{.eps, .jpg,.ps}
%\newcommand{\picspace}{\vspace{0in}}
%\newcommand{\picspacehalf}{\vspace{0in}}
%\fi
%\ifpdf
%\DeclareGraphicsExtensions{.pdf, .jpg}
%\else
%\DeclareGraphicsExtensions{.eps, .jpg,.ps}
%\fi
%%%%%%%%%%%%%%%%%%%%%%%%%%%%%%%%%%%%%%%%%%

%%%%%%%%%%%%%%%%%%%%%%%%%%%%%%%%%%%%%%%%%%
%Define Title, Author, Address, Preprint#

%\preprint{\vbox{ \hbox{}   \hbox{} }}

\title{QED Contribution to the Color-Singlet $\jpsi$ Production in $\Upsilon$ Decay Near the Endpoint}
	
\author{Xiaohui Liu}
\affiliation{Department of Physics and Astronomy,
	University of Pittsburgh,
        Pittsburgh, PA 15260\vspace{0.2cm}}

\date{\today\\ \vspace{1cm} }

%%%%%%%%%%%%%%%%%%%%%%%%%%%%%%%%%%%%%%%%%%

%%%%%%%%%%%%%%%%%%%%%%%%%%%%%%%%%%%%%%%%%%
%Create the title page

\begin{abstract}

A recent study indicates that the $\alpha^2\alpha_s^2$ order QED processes of
$\Upsilon \to \jpsi + X$ decay are 
compatible with those of QCD processes. However, in the endpoint region, the 
Non-relativistic QED (NRQED) calculation breaks down since the collinear degrees of freedom are
missing under the framework of this effective theory.  
 In this paper we 
apply the soft collinear effective theory (SCET) to study the color-singlet QED process at the kinematic limit. Within this approach we are
able to sum the kinematic logarithms by running operators using the renormalization group equations of SCET, which will lead to a dramatic change in the 
momentum distribution near the endpoint and the spectrum shape  
consistent with the experimental results. 
\end{abstract}

\maketitle

%%%%%%%%%%%%%%%%%%%%%%%%%%%%%%%%%%%%%%%%%%
%\tighten
\newpage
%%%%%%%%%%%%%%%%%%%%%%%%%%%%%%%%%%%%%%%%%%
%Main body of the paper

\section{Introduction}

During the past 15 years, the interactions of non-relativistic heavy quarks inside  
quarkonium have been
understood to some extent using the framework of non-relativistic effective
theories~\cite{Bodwin:1995prd,Luke:2000prd}. These theories 
reproduce the physics of full QCD or QED by adding local 
interactions that 
systematically incorporate relativistic corrections through any given order 
in the heavy quark velocity $v$.  They provide 
generalized 
factorization theorems that include nonperturbative
corrections to the color-singlet model, including color-octet decay mechanisms.
 All infrared divergences can be factored into nonperturbative matrix elements,
 so that infrared safe calculations of inclusive decay rates are 
possible~\cite{Bodwin:1992prd}.
These non-relativistic effective theories solve some important 
phenomenological problems in quarkonium physics. 
For instance, they provide the most convincing explanation to the 
surplus $\jpsi$ and $\psi'$ production at the Tevatron~\cite{Braaten:1995prl},
in which a gluon fragments into a color-octet $c\bar{c}$ pair in a pointlike
color-octet S-wave state which evolves nonperturbatively into the charmonium
states plus light hadrons. The factorization formalism allows these 
fragmentation procedures to be factored into the product of short distance 
coefficients and long distance matrix elements among which 
the leading one is  
$\langle {\cal O}^{\bf 8}_{\psi(\psi')}[{}^3S_1]\rangle$ where 
$ {\cal O}^{\bf 8}_{\psi(\psi')}$ are local four-fermon operators in terms of 
the non-relativistic fields.

There are, however, some problems that remain to be solved. 
One challenging problem is with the polarization of $\jpsi$ 
at the Tevatron. The same mechanism that produces the $\jpsi$ described above 
predicts the $\jpsi$ should become transversely 
polarized as the transverse momentum $p_\perp$ becomes 
much larger than $2m_c$~\cite{Cho:1995plb}. 
Though the theoretical prediction is consistent with the experimental data at 
intermediate $p_\perp$, at the largest measured values of $p_\perp$ the 
$\jpsi$ is observed to be slightly longitudinally polarized and 
discrepancies at the $3\sigma$ level are seen in both 
prompt $\jpsi$ and $\psi'$ polarization measurements~\cite{Affolder:2000prl}.

A new problem arose as a result of measurements of the 
spectrum of $\jpsi$ produced in the $\Upsilon(1S)$ decay by the CLEO III 
detector at CESR~\cite{Briere:2004prd}. NRQCD calculations have been made
for the production of $\jpsi$ through both color-singlet and color-octet 
configurations~\cite{Li:2000plb,Cheung:1996prd}. Theoretical calculations
predict that the color-singlet process $\Upsilon(1S)\to \jpsi c\bar{c}g+X$
features a soft momentum spectrum.
% which peaks near $2.2{\rm GeV}$ and the $\jpsi$ momentum has a kinematic limit $3.3{\rm GeV}$~\cite{Li:2000plb}. 
Meanwhile, the theoretical estimates based on color-octet contributions 
indicate that the momentum spectrum peaks near the kinematic 
endpoint~\cite{Cheung:1996prd}. In contrast to the theoretical predictoins,
the experimentally measured momentum spectrum is significantly softer than
predicted by the color-octet model and somewhat softer than the color-singlet
case~\cite{Briere:2004prd}.    

A more detailed study on the color-singlet contribution to this process
has been presented recently~\cite{He:arXiv0911}. It was found that the 
contribution of the 
color-singlet QED process is comparable with the QCD process.
NRQED calculations indicate that the QED process will
give a large contribution to the spectrum near the end-point that is not observed in 
the data.  This contribution results from the $J/\psi$ being produced back-to-back with a pair of gluons forming a low-mass jet.  However, in this region of phase space, the NRQED calculation breaks down, since it
does not contain the correct degrees of freedom.  NRQED contains soft
quarks, photons and gluons, but it does not contain quarks and gluons
moving collinearly.
The correct effective theory to use in situations where there is both soft 
and collinear physics is 
Soft-Collinear Effective 
Theory (SCET)~\cite{BauerO:2001prd,BauerS:2001prd,Bauer:2001plb,Bauer:2002prd}.  
%to study the effects systematically~\cite{Briere:2004prd} 

A similar situation happens when studying $e^+e^-\to \jpsi +X$. The combination of SCET and NRQCD has been successful in reproducing the shape of the 
measured $\jpsi$ momentum spectrum in 
$e^+e^-\to \jpsi +X$~\cite{FlemingIntro:2003prd}. SCET has the power to 
describe the endpoint regime by including the light energetic degrees of 
freedom. In addition, the renormalization group equations of SCET can be used to
resum large perturbative logarithmatic correctoins. THE nonperturbative NRQCD martix elements
arise naturely in deriving the factorizatoin theorem using SCET.

In this paper, we use SCET to study the color-singlet 
contribution to the $\Upsilon \to \jpsi + X$ decay near the endpoint via
a virtual photon. 
We derive the factorization theorm in SCET for this process.
We find that the spectrum is softer than the tree order prediction
of NRQED 
when including perturbative 
and nonperturbative corrections near the 
endpoint, giving better agreement with the data than the previous 
predictions.

\section{Matching and Factorization}
In this section, we derive the SCET factorization theorem for the  
color-singlet contribution to $\Upsilon \to \jpsi + X$ via a virtual photon
near end-point.
This factorization formula is crucial since the NRQED does not properly
include the relevent collinear degrees of freedom and thus breaks down in
this regime. This can be understood by analyzing the kinematics near
the end-point. In the centre-of-mass (COM) frame, we have
\bea
&&\pup^\mu = \,
\frac{\mup}{2}n^\mu + \frac{\mup}{2}\bn^\mu + k^\mu_\Upsilon\,, \nn \\
&&p_\psi^\mu = \,
\frac{\mpsi^2}{2z\mup}n^\mu + \frac{z\mup}{2}\bn^\mu +k^\mu_\psi\,, \nn \\
&&p_X^\mu = \,
\frac{\mup}{2}\left[\left(1-\frac{r}{z}\right)n^\mu + \,
(1-z)\bn^\mu \right] + k_X^\mu \,.
\eea
Here $n=(1,0,0,1)$ and $\bn = (1,0,0,-1)$, we 
have defined $z= (E_\psi+ p_\psi)/\mup$ and $r = m_c^2/m_b^2$.
We also assumed
that $M_\psi = 2 m_c$ and $\mup = 2m_b$. 
$k_\Upsilon^\mu$ and $k_\psi^\mu$ are the residual momentum 
of the $Q\bar Q$ pair inside the $\Upsilon$ and $\jpsi$ respectively. 
Near the kinematic 
endpoint, the variable $z \to 1$ and thus the jet invariant mass 
approaches zero.
In NRQED, an expansion of $k^\mu/m_X$ is performed and hence the 
jet mode is integrated out, which is only
valid when the jet mass is large compare to the residual momentum.  
The invariant mass of the jet is large away from the endpoint.  
As $z \to 1$, the jet becomes energetic, with small invariant mass.
Hence we must
keep $k^\mu/m_X$ to all orders. As a result, the standard NRQED 
factorization breaks down at the endpoint. SCET is the appropriate framework 
for properly including the collinear modes needed in the endpoint  
in order to make reasonable predictions.

To derive the factorization theorem in SCET, we start with 
the optical theorem in which the decay rate is written as
\bea\label{Optic}
2E_\psi \frac{\mathrm{d}\Gamma}{\mathrm{d}^3 p_\psi} \,
=\frac{1}{32\pi^3 m_b} \sum_X \int \mathrm{d}^4 y\, e^{-iq\cdot y}\,
\langle \Upsilon | {\cal O}^\dagger (y) |\jpsi+X\rangle \,
\langle \jpsi +X | {\cal O}(0) | \Upsilon \rangle \,,
\eea
where the summation includes integration over the $X$ phase space, 
which includes both the ultrasoft (usoft) $X_u$ and collinear 
$X_c$ sectors. The SCET operator ${\cal O}$ is of the form
\bea\label{ScetOpt}
{\cal O} = \sum_\omega e^{-i(\mup v + \bnP \frac{n}{2})\cdot y }\,
C(\mu,\omega)\,
\Gamma_{\alpha \beta \mu \nu}\,
 {\cal J}^{\alpha \beta}(\omega)\,
{\cal O}_{\jpsi}^\mu {\cal O}_\Upsilon^\nu \,,
\eea
where the Wilson coefficient $C(\mu,\omega)$ is obtained by matching from 
QCD to SCET at some hard
scale $\mu = \mu_H$. The operator is contrained by the gauge invariance. 
In our case, to leading order we have 
\bea
{\cal J}^{\alpha\beta}(\omega) &=& Tr[B_{\perp\omega_1}^\alpha 
B^\beta_{\perp\omega_2}] \,, \\
{\cal O}_{\jpsi}^\mu &=& \psi^\dagger_{\bar c}\,
\left( \Lambda_1\cdot \sigma \right)^\mu \chi_c \,, \\
{\cal O}_{\Upsilon}^\nu &=& \chi^\dagger_{\bar b}\,
\left( \Lambda_2\cdot \sigma \right)^\nu \psi_b \,. 
\eea
Here the $\Lambda$'s boost the $\jpsi$ or $\Upsilon$ from the COM frame to an arbitrary
frame. $\psi$ and $\chi$ are the heavy quark and 
antiquark fields which
create or annihilate the constituent heavy (anti-)quarks inside the quarkonia. 
The collinear
gauge invariant field strength $B^\alpha_\perp$ is built out of the 
collinear gauge field $A^\alpha_{n,q}$
\bea
B_\perp^\alpha = \frac{-i}{g_s}W^\dagger_n\,
\left( \cP^\alpha_\perp + g_s(A^\alpha_{n,q})_\perp \right) W_n \,.
\eea
where
\bea
W_n = \sum_{\rm perms} \exp\left(-g_s\frac{1}{\bnP}\bn\cdot A_{n,q} \right)\,
\eea
is the collinear Wilson line. The operator $\cP$ is used to project out the 
large momentum label~\cite{Bauer:2001plb}.

\begin{figure}[t]
\begin{center}
\includegraphics[width=12cm]{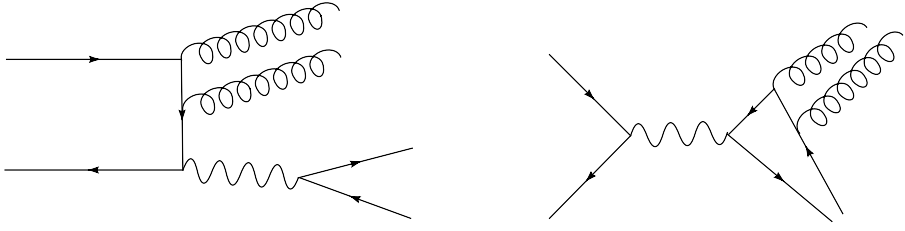}
\caption{\small 
Diagrams for the QED contribution to the color-singlet $\jpsi$ production
via $\Upsilon$ decay at order $\alpha^2\alpha_s^2$. 
\label{MatchQED}}
\end{center}
\end{figure} 

The hard coefficient containing the spin structure 
is obtained by matching the Feynman diagrams shown in 
fig.~\ref{MatchQED}, which gives
\bea
\Gamma_{\alpha\beta\mu\nu} =i \frac{32\pi^2}{2N_c}\,
\frac{e_c e_b \alpha \alpha_s}{m_c m_b} g^\perp_{\alpha \beta}\,
 g^\perp_{\mu\nu}\,,
\eea
where $g^\perp_{\mu\nu} = g_{\mu\nu}-(n_\mu\bn_\nu+n_\nu \bn_\mu)/2$.
We have chosen the hard coefficient so that the Wilson coefficient
$C(\mu,\omega)$ is $1$ at the hard scale $\mu_H$.

Inserting the operator in Eq.~(\ref{ScetOpt}) into Eq.~(\ref{Optic}),
${\cal O}^\dagger(y)$ picks an additional phase, and the differential
rate becomes
\bea\label{MoOptic}
2E_\psi \frac{\mathrm{d}\Gamma}{\mathrm{d}^3 p_\psi} \,
&=&\frac{1}{32\pi^3 m_b}\,
\sum_X \sum_{\omega\omega'}C^\dagger(\mu,\omega) C(\mu,\omega')\,
\int \mathrm{d}^4 y\, e^{-i\mup/2(1-z)\bn\cdot y}\,
\Gamma^\dagger_{\alpha\beta\mu\nu}
\Gamma_{\alpha'\beta'\mu'\nu'} \nn \\ 
&&\times \langle \Upsilon | {\cal J}^{\alpha\beta\dagger}_\omega\,
{\cal O}_{\jpsi}^{\mu \dagger}
{\cal O}_\Upsilon^{\nu \dagger}(y) |\jpsi+X\rangle \,
\langle \jpsi +X | {\cal J}^{\alpha'\beta'}_{\omega'}\,
{\cal O}^{\mu'}_{\jpsi}\,
{\cal O}^{\nu'}_{\Upsilon}(0) | \Upsilon \rangle \nn \\
&\equiv& \sum_{\omega\omega'}C^\dagger(\mu,\omega) C(\mu,\omega')\,
\Gamma_{\alpha\beta\mu\nu}^{\dagger}\,
\Gamma_{\alpha'\beta'\mu'\nu'} \,
{\cal A}^{\alpha\beta\mu\nu,\alpha'\beta'\mu'\nu'}_{\omega\omega'} \,,
\eea
In the exponent of Eq.~(\ref{MoOptic}), we have used 
$q^\mu - \mup v^\mu + \bnP n^\mu/2 \approx \mup/2(1 - z)\bn^\mu$.
Furthermore, we can decouple the usoft modes from the collinear degrees of
freedom using the field redefinition~\cite{Bauer:2002prd} 
\bea
A_{n,q}^\mu = Y A_{n,q}^{(0)\mu} Y^\dagger \,.
\eea
The fields with the superscript $(0)$ do not interact with usoft degrees
of freedom. In the color-singlet contribution the usoft Wilson lines $Y$ cancel
since $Y^\dagger Y = 1$. The $\Upsilon$ and the $\jpsi$ states contain
no collinear quanta, so we can write
\bea
{\cal A}^{\alpha\beta\mu\nu,\alpha'\beta'\mu'\nu'}_{\omega\omega'} \,
&=& \frac{1}{32\pi^3 m_b}\,
\int \mathrm{d}^4 y\, e^{-i\mup/2(1-z)\bn\cdot y} \nn \\ 
&&\times \langle \Upsilon |
{\cal O}_{\jpsi}^{\mu \dagger}
{\cal O}_\Upsilon^{\nu \dagger}(y) \,
a_\psi^\dagger a_\psi
{\cal O}^{\mu'}_{\jpsi}\,
{\cal O}^{\nu'}_{\Upsilon}(0) | \Upsilon \rangle \,
\langle 0|{\cal J}^{\alpha \beta \dagger}_\omega(y)\,
{\cal J}^{\alpha'\beta'}_{\omega'}(0) |0\rangle \,.
\eea
Here we defined an interpolating field, $a_\psi$, for the $\jpsi$ and used
the completeness of states in the usoft and collinear fields
\bea
\sum_{X_u}|\jpsi+X_u \rangle \langle \jpsi+X_u| &=& \,
|\jpsi \rangle \langle \jpsi | \equiv a^\dagger_\psi a_\psi \,, \\
\sum_{X_c} | X_c \rangle \langle X_c| &=&  1 \,. 
\eea

The $\Upsilon$ is a
very compact bound state, due to the large b-quark mass. 
In a multipole expansion, long
wavelength gluons interacts with the 
$\Upsilon$ color charge distribution through its 
color dipole moment since the state itself is color neutral. 
In the theoretical limit of very heavy bottom
quark, this coupling to the dipole vanishes~\cite{Bobeth:2008prd}. 
Therefore we are able to write
\bea
{\cal A}^{\alpha\beta\mu\nu,\alpha'\beta'\mu'\nu'}_{\omega\omega'} \,
&=& \frac{1}{32\pi^3 m_b}\,
 \int \mathrm{d}^4 y\, e^{-i\mup/2(1-z)\bn\cdot y} \nn \\ 
&&\hspace{-7.ex}\times \langle \Upsilon |
{\cal O}_\Upsilon^{\nu \dagger}(y) \,
{\cal O}^{\nu'}_{\Upsilon}(0) | \Upsilon \rangle \,
\langle 0 | {\cal O}_{\jpsi}^{\mu \dagger}(y)\,
a_\psi^\dagger a_\psi
{\cal O}^{\mu'}_{\jpsi}(0)\,
|0\rangle \,
\langle 0|{\cal J}^{\alpha \beta \dagger}_\omega(y)\,
{\cal J}^{\alpha'\beta'}_{\omega'}(0) |0\rangle \,.
\eea

To proceed, we introduce the shape function for $\jpsi$
\bea\label{PsiShape}
S_{\psi}(l^+) = \int \frac{\mathrm{d}y^-}{4\pi}\,
e^{-\frac{i}{2}l^+y^-}\,
\frac{\langle 0|\left[\,
\chi_{\bar c}^\dagger \sigma_i \,
\psi_c(y^-) \,
a^\dagger_\psi a_\psi \,
\psi_c^\dagger \sigma_i \,
\chi_{\bar c} \,
\right]|0\rangle}{4m_c\langle {\cal O}^{\bf 1}_\psi[{}^3S_1]\rangle}\,,
\eea
as well as the shape function for $\Upsilon$
\bea
S_{\Upsilon}(l^+) =  \int \frac{\mathrm{d}y^-}{4\pi}\,
e^{-\frac{i}{2}l^+y^-}\,
 \frac{\langle\Upsilon|\chi_{\bar b}^\dagger \sigma_i \,
\psi_b(y^-)\psi_b^\dagger \sigma_i \chi_{\bar b}|\Upsilon\rangle}{4m_b\langle\Upsilon| {\cal O}^{\bf 1}_\Upsilon[{}^3S_1]|\Upsilon\rangle}\,.
\eea
 Both shape functions are normalized so that 
$\int \mathrm{d}l^+S_{\psi,\Upsilon}(l^+) = 1$.  The are color-singlet shape functions can be related simply to the color-singlet NRQCD matrix elements~\cite{Rothstein:1997plb},
\bea
\langle \chi^\dagger \sigma_i \,
\psi\,
\delta(i n\cdot \partial - k^+)  \,
\psi^\dagger \sigma_i \,
\chi \rangle \,
=\Theta(k^+) \langle \chi^\dagger \sigma_i \,
\psi\,
\psi^\dagger \sigma_i \,
\chi \rangle \,,
\eea
which amounts to a shift from the partonic to hadronic endpoint.

In addition a jet function $J_\omega(k^+)$ is defined as
\bea
&&\langle 0| Tr\left[ B_{\alpha}^{\perp}B_{\beta}^{\perp}\right](y)\,
 Tr\left[ B_{\alpha'}^{\perp}B_{\beta'}^{\perp}\right](0) |0 \rangle \nn\\
&=& i \frac{N_c^2-1}{2}\,
(g_{\alpha\alpha'}g_{\beta\beta'}\,
+ g_{\alpha\beta'}g_{\beta\alpha'})\,
\delta_{\omega\omega'}\,
\int \frac{\mathrm{d}k^+}{2\pi} \delta^{(2)}(y^\perp )\,
\delta(y^+) e^{-\frac{i}{2}k^+y^-} J_\omega(k^+) \,.
\eea
The leading order result for the collinear jet function is~\cite{Fleming:2003prd}
\bea
J_\omega(k^+) = \frac{1}{8\pi}\Theta(k^+) \int_0^1 \mathrm{d}\xi\,
\delta_{\xi,(\mup+\omega)/(2\mup)} \,.
\eea
Using the spin symmetry relation~\cite{Braaten:1996prd} 
\bea
&&\Lambda_i^{\delta}\Lambda_j^{\delta'} \langle \dots {\bf \sigma}^i \dots \,
{\bf \sigma}^j \dots \rangle = \nn \\
&&\hspace{10.ex} \frac{1}{3}\delta^{ij}\Lambda_i^{\delta}\Lambda_j^{\delta'} \,
\langle \dots {\bf \sigma}^k \dots \,
{\bf \sigma}^k \dots \rangle \,,
\eea
and applying the
identity $\delta^{ij}\Lambda_i^{\delta}\Lambda_j^{\delta'} 
= (v^{\delta}v^{\delta'}-g^{\delta \delta'})$, where $v^\delta$ is 
the four-velocity
of the $\Upsilon$ or $\jpsi$, we can write the decay rate as
\bea
\frac{\mathrm{d}\Gamma}{\mathrm{d}p_\psi} = \,
\Gamma_0 P[x,r] \int_{-1}^1 \frac{\mathrm{d}\xi}{2} |C(\mup \xi,\mu)|^2 \,
\Theta(\mup -2 E_X) \,,
\eea
in which
\bea
\Gamma_0 = \frac{4\pi}{9}\frac{N_c^2-1}{N_c^2}\,
\frac{e_b^2e_c^2\alpha^2\alpha_s^2}{m_b^3 m_c^3} \,
\frac{(1-r)^2}{1+r}
\langle\Upsilon| {\cal O}^{\bf 1}_\Upsilon[{}^3S_1]|\Upsilon\rangle \,
\langle {\cal O}^{\bf 1}_{\psi}[{}^3S_1]\rangle \,,
\eea
and $P[x,r] = (x^2-4r)(1+r)/(x(1-r)^2)$. Near the end-point, $P[x,r] \to 1$.
The variable $x$ is defined as $x = E_\psi/m_b$.
It is straight forward to check that to the leading order
the differential decay rate reproduces the NRQED calculation~\cite{He:arXiv0911}.

\section{Resumming Sudakov Logarithms and Phenomenology}

SCET has the power to sum logarithms using the renormalization group
 equations (RGEs). Large logarithms arise
naturally in the processes involving several well-separated scales and 
will cause the perturbative expansion breaking down.
By matching onto an effective theory, the large scale is removed and
replaced by a running scale $\mu$. After matching at the high scale, 
the operators are run to
the low scale using the RGEs. 
This sums all large logarithms into an overall factor, and any
logarithms that arise in the perturbative expansion of the effective theory 
are of order one.

In the previous section, we have matched onto the SCET color-singlet 
operator, by
intergrating out the large scale $\mu_H$, replacing it with a running 
scale $\mu$. We now run the
operator from the hard scale to the collinear scale,
 which sums all logarithms. The counterterm as well as the anomalous dimension 
used for running the operator in the RGEs have already been calculated in 
Ref.~\cite{Fleming:2003prd} , 
and we 
can lift the results from that paper. The result for the resummed differential 
decay rate is given by
\bea\label{resumrate}
\frac{1}{\Gamma_0}\frac{\mathrm{d}\Gamma_{\rm resum}}{\mathrm{d} p_\psi} =\,
P[x,r] \Theta(\mup - 2 E_X) \int \mathrm{d}\eta \,
\left[\frac{\alpha_s(\mu_c)}{\alpha_s(\mu_H)} \right]^{2\gamma(\eta)} \,,
\eea
where $\gamma$ is defined as
\bea
\gamma \equiv \frac{2}{\beta_0}\left[  \,
C_A\left( \frac{11}{6}+ \left(\eta^2 +(1-\eta)^2\,
\right)\left(\frac{1}{1-\eta}\ln \eta +\frac{1}{\eta}\ln(1-\eta)  \,
\right)\,
\right)-\frac{n_f}{3}\,
\right]\,.
\eea
To sum the large logarithms, the collinear scale $\mu_c^2$ is chosen to be 
approximately $m_X^2$ and the hard scale is set to be $\mu_H = 2m_c$, in
same way as in Ref.~\cite{He:arXiv0911}.

The result from Eq.~(\ref{resumrate}) sums up the leading logarithmic
corrections which are important only near the endpoint. 
Away from the endpoint, the
logarithms that we have summed are not important and contributions 
that we neglected in
the endpoint become dominant. We therefore would like to interpolate between 
the leading
order NRQED color-singlet calculation away from the endpoint and 
the resummed result in the
endpoint. To do this, we define the interpolated differential rate as
\bea\label{inter}
\frac{1}{\Gamma_0} \frac{\mathrm{d}\Gamma}{\mathrm{d}p_\psi} =\,
\left(\frac{1}{\Gamma_0} \,
\frac{\mathrm{d}\Gamma^{\rm dir}_{\rm LO}}{\mathrm{d}p_\psi}-P[x,r]\,
\right) + \,
\frac{1}{\Gamma_0} \frac{\mathrm{d}\Gamma_{\rm resum}}{\mathrm{d}p_\psi} \,.
\eea
The first term in parentheses vanishes when approaching the kinematic limit,
leaving only the resummed contribution in that region. Away from
the endpoint the resummed contribution combines with the $-P[x,r]$ to
give higher order corrections in $\alpha_s(\mu_H)$ to the spectrum.

\begin{figure}[t]
\begin{center}
\includegraphics[width=12cm]{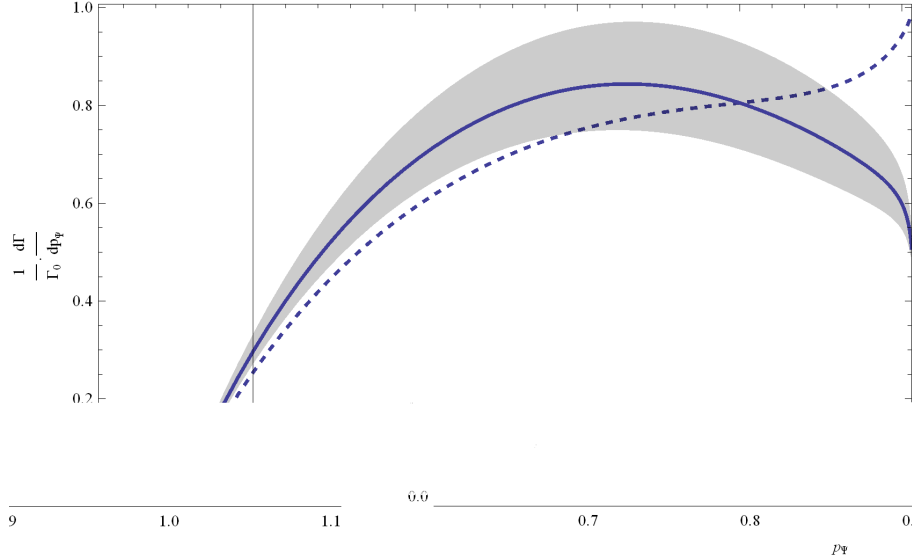}
\caption{\small 
The decay rate $1/\Gamma_0 \mathrm{d}\Gamma/\mathrm{d}p_\psi$ 
via QED process. The dashed curve is the tree level direct
rate~\cite{He:arXiv0911}. The solid line presents the interpolated 
resummed direct rate. The 
shaded band is obtained by varying the collinear scale from 
$\mu_c = m_X/\sqrt{2}$ to $\mu_c = \sqrt{2}m_X$, since the choice of scale
could only be determined by a higher order calculation.
\label{HarCorr}}
\end{center}
\end{figure}

In fig.~\ref{HarCorr}, we compare the
resummed, interpolated decay rate, Eq.~(\ref{inter}), 
to the leading-order color-singlet result~\cite{He:arXiv0911}.
We use $m_c = 1.548{\rm GeV}$ and $m_b = 4.73{\rm GeV}$. $\Lambda_{\rm QCD}$
is set to $0.21{\rm GeV}$ so that  $\alpha_s(2m_c) = 0.259$.
In our figure, the dashed
line presents the leading-order color-singlet calculation and the solid
curve corresponds to the interpolated decay rate with the collinear
scale chosen as $\mu_c = m_X$. The 
shaded band is obtained by varying the collinear scale from 
$\mu_c = m_X/\sqrt{2}$ to $\mu_c = \sqrt{2}m_X$, since the choice of scale
could only be determined by higher order corrections.
After resumming the spectrum shape softens near the end-point and is thus more
consistent with experimental data~\cite{Briere:2004prd}.

\section{Conclusion}
In this work, we study the color-singlet QED process for $\jpsi$ production
in $\Upsilon$ decay in the kinematic limit region. Since the NRQED breaks
down ate this limit, we apply the SCET to study the spectrum. 
Our calculation
consists of matching onto a color-singlet operator in SCET by 
integrating out the hard scale. 
Once the usoft modes are decoupled from the collinear modes 
using a field redefinition, we
are able to show a factorization theorem for the differential decay rate, 
in which the decay 
rate can be factorized into a hard piece, a collinear jet function, 
and usoft functions. As
pointed out by Ref.~\cite{Rothstein:1997plb} 
the usoft function in this case can be calculated, resulting in just a
shift from the partonic to the physical endpoint.

By running the resulting rate from the hard scale $\mu_H$ 
to the collinear scale $\mu_c$, we sum the large Sudakov
logarithms. Finally, we combine the SCET calculation with the leading order,
color-singlet NRQED result to make a prediction for the color-singlet 
contribution via QED process to the
differential decay rate spectrum over the entire allowed kinematic range.

\section*{Acknowledgments}

I would like to thank  Professor A.~K.~Leibovich for guidances and carefully 
reading the manuscript and checking all the calculations. XL was supported 
in part by the National Science Foundation under Grant No. PHY-0546143.

\end{document}